\documentclass[12pt]{spieman}  
\usepackage{amsmath,amsfonts,amssymb}
\usepackage{graphicx}
\usepackage{setspace}
\usepackage{tocloft}
\usepackage{lineno}

\title{Demonstrating repetitive non-destructive readout (RNDR) with SiSeRO devices}

\author[a]{Tanmoy Chattopadhyay}
\author[a]{Sven Herrmann}
\author[a]{Peter Orel}
\author[b]{Kevan Donlon}
\author[c]{Gregory Prigozhin}
\author[a]{R. Glenn Morris}
\author[b]{Michael Cooper}
\author[c]{Beverly LaMarr}
\author[c]{Andrew Malonis}
\author[a,d,e]{Steven W. Allen}
\author[c]{Marshall W. Bautz}
\author[b]{Chris Leitz}

\affil[a]{Kavli Institute of Astrophysics and Cosmology, Stanford University, 452 Lomita Mall, Stanford, CA 94305, USA}
\affil[b]{MIT Lincoln Laboratory, Lexington, MA, USA}
\affil[c]{Kavli Institute for Astrophysics and Space Research, Massachusetts Institute of Technology, Cambridge, MA, USA}
\affil[d]{SLAC National Accelerator Laboratory, 2575 Sand Hill Road, Menlo Park, CA 94025, USA}
\affil[e]{Department of Physics, Stanford University, 382 Via Pueblo Mall, Stanford CA 94305, USA}

\cftpagenumbersoff{figure}
\cftpagenumbersoff{table}

\begin{document} 
\maketitle

\begin{abstract}
We demonstrate so-called repetitive non-destructive readout (RNDR) for the first time on a Single electron Sensitive Readout (SiSeRO) device. SiSeRO is a novel on-chip charge detector output stage for charge-coupled device (CCD) image sensors, developed at MIT Lincoln Laboratory. This technology uses a p-MOSFET transistor with a depleted internal gate beneath the transistor channel. The transistor source-drain current is modulated by the transfer of charge into the internal gate. RNDR was realized by transferring the signal charge non-destructively between the internal gate and the summing well (SW), which is the last serial register. The advantage of the non-destructive charge transfer is that the signal charge for each pixel can be measured at the end of each transfer cycle and by averaging for a large number of measurements ($\mathrm{N_{cycle}}$), the total noise can be reduced by a factor of 1/$\mathrm{\sqrt{N_{cycle}}}$. In our experiments with a prototype SiSeRO device, we implemented nine ($\mathrm{N_{cycle}}$ = 9) RNDR cycles, achieving around 2 electron readout noise (equivalent noise charge or ENC) with spectral resolution close to the fano limit for silicon at 5.9 keV. These first results are extremely encouraging, demonstrating successful implementation of the RNDR technique in SiSeROs. They also lay foundation for future experiments with more optimized test stands (better temperature control, larger number of RNDR cycles, RNDR-optimized SiSeRO devices) which should be capable of achieving sub-electron noise sensitivities. This new device class presents an exciting technology for next generation astronomical X-ray telescopes requiring very low-noise spectroscopic imagers. The sub-electron sensitivity also adds the capability to conduct in-situ absolute calibration, enabling unprecedented characterization of the low energy instrument response.  
\end{abstract}


{\noindent \footnotesize\textbf{*}Tanmoy Chattopadhyay,  \linkable{tanmoyc@stanford.edu} }

\begin{spacing}{1}   


\section{INTRODUCTION}
\label{sec:intro}  
The concept  of Repetitive non-destructive readout (RNDR) exists for sometime now (see Janesick et al. 1990 \cite{janesick90} in the context of skipper CCDs and Kraft et al. 1995 \cite{KRAFT95}) which allows to non-destructively measure charge contained in a pixel multiple times using floating gate amplifiers. They achieved a readout noise better than the noise of a single amplifying stage.
In recent times, this technique has seen progress in various silicon sensor technologies, making them sensitive to single
electrons, e.g. skipper CCDs \cite{tiffenberg17,rodrigues21}, Depleted P-channel Field-Effect Transistor (DEPFET \cite{wolfel06,treberspurg22_rndrdepfet}) and proposed Complementary Metal-Oxide-Semiconductor (CMOS) imaging sensors (CIS) with RNDR \cite{Stefanov2020_skippercmos}. 
The final noise in RNDR measurements is given by $\mathrm{\sigma_N = \sigma/\sqrt{N_{cycle}}}$, where $\mathrm{\sigma}$ is the noise for a single read and $\mathrm{N_{cycle}}$ is the number of repetitive cycles. The final noise, therefore, depends on the number of RNDR cycles, the single readout noise, and thermal dark current (noise). There could be other spurious noise sources like clocking the charge back and forth can introduce clock induced charge or potential light glow from the amplifiers adds to the thermal noise floor. These noise sources are particularly important at cryogenic temperatures when the thermal dark noise is low.
The Single electron Sensitive Read Out (SiSeRO) readout stage developed by MIT Lincoln Laboratory (MIT-LL) is a novel charge detector for X-ray CCDs, with a working principle similar in some respects to DEPFET sensors \cite{kemmer87_depfet,strueder00_depfet_imager} and extremely high responsivity floating-gate amplifiers \cite{matsunaga91}. By transferring the signal charge from and to the buried gate, SiSeRO devices can operate in RNDR mode. The advantage of this specific technology is that it can be used not only to augment CCDs but also to build active pixel sensor (APS) arrays in which each pixel includes a SiSeRO, thereby enabling extremely low noise performance at high readout speeds, without the large distance charge transfer needed for CCDs. 

\begin{figure}[ht!]
    \centering
   \includegraphics[width=1\linewidth]{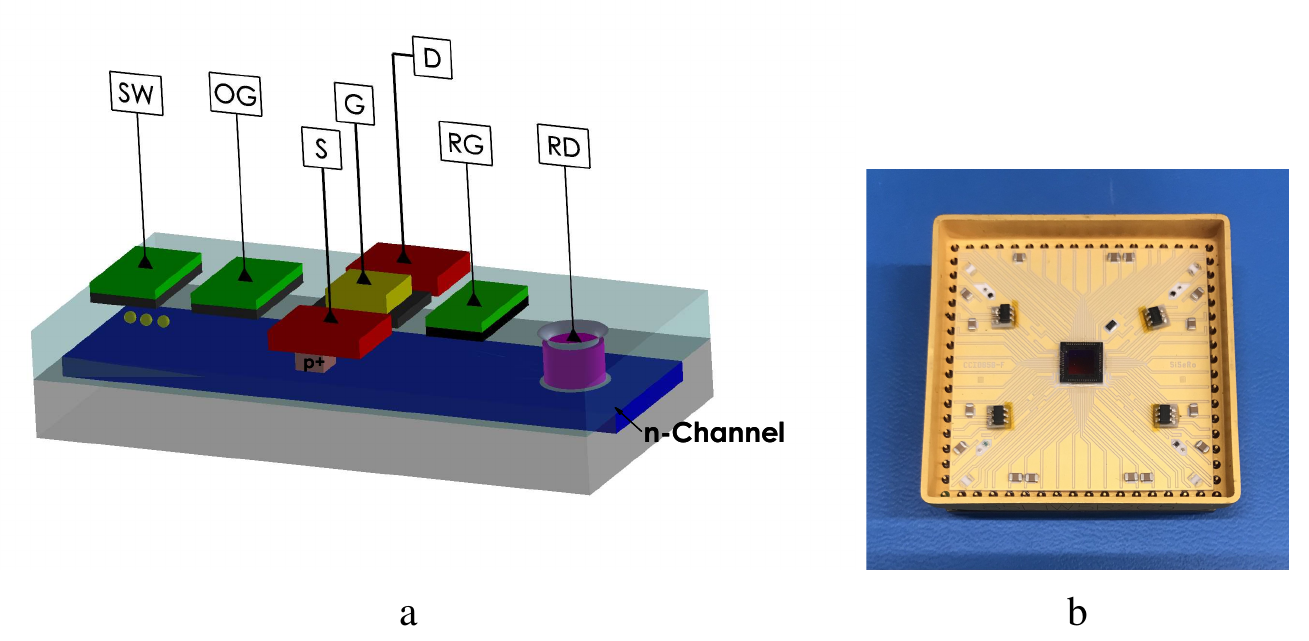}
    \caption{(a) Schematic of a SiSeRO output stage, utilizing a p-MOSFET transistor with an internal gate beneath the p-MOSFET. The internal channel is emptied or reset using the reset transistor (RG) and reset drain (RD). Charge transferred from the last serial gate (summing well or SW) to the internal gate and through the output gate (OG) modulates the drain current of the p-MOSFET. (b) A prototype SiSeRO device (CCID-93) in its package, used for testing. 
    }
    \label{fig:sisero}
\end{figure}

The devices studied here consist of a CCD pixel array with a SiSeRO amplifier straddling the n-channel of the CCD’s output register (see the simplified 3D schematic in Fig. \ref{fig:sisero}a). When a charge packet is present in the CCD channel beneath the SiSeRO p-MOSFET, it modulates the transistor drain current, which is then readout by the readout electronics \cite{chattopadhyay22_sisero,Chattopadhyayetal2022}.
\begin{figure}[ht!]
    \centering
   \includegraphics[width=.5\linewidth]{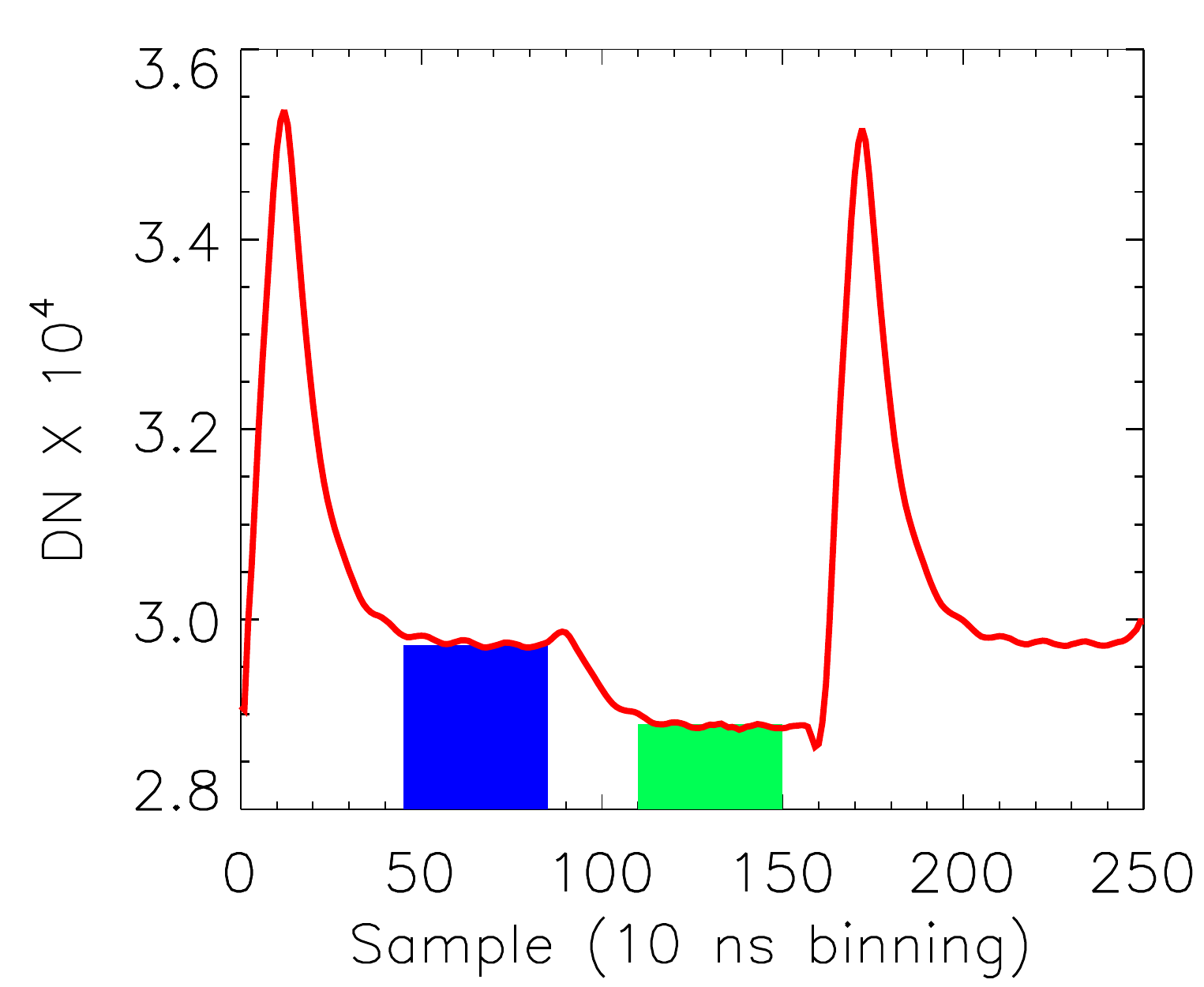}
    \caption{SiSeRO output video waveform. The shaded regions represent the selected baseline (blue) and signal samples (green) respectively. The difference between the two levels is proportional to the signal strength.}
    \label{fig:sisero_wave}
\end{figure}
For a single polysilicon MOSFET in SiSeROs, the internal gate minimizes parasitic capacitance on the sense node, resulting in a high conversion gain and minimized noise. 
Recently, for a prototype CCD device (Fig. \ref{fig:sisero}b)  with a buried-channel SiSeRO at the output stage, a read noise of 6 $\mathrm{e}^{-}_{\mathrm{RMS}}$ was achieved. This was further improved to 4.5 $\mathrm{e}^{-}_{\mathrm{RMS}}$ by applying digital filtering techniques to minimize the 1/f noise \cite{chattopadhyay23_sisero}. We measured a gain or responsivity of 800 pico-ampere (pA) per electron for this device. The FWHM of the 5.9 keV Mn K$_\alpha$ line was measured to be $\sim$ 132 eV. The test device, from the CCID-93 family developed by MIT-LL, has an active area of $\sim$4 mm $\times$ 4 mm with a 512 $\times$ 512 array of 8 $\mu$m pixels. 
Figure \ref{fig:sisero_wave} shows a typical SiSeRO digital video waveform at $\sim$625 kpixel/s readout rate (each pixel is $\sim$1600 ns long) sampled every 10 ns. 
The change from the baseline (the shaded blue region) to the signal level (the shaded green region) after the charge packet is transferred is proportional to the signal strength. The signal amplitude is extracted by taking the difference between these two levels (correlated double sampling or CDS) for each pixel, which is then used to generate the 2D images. 

Since the charge packet in the internal gate is unaffected by the readout process, SiSeROs offer the potential for repetitive non-destructive readout (RNDR). The charge can be moved around like any charge packet in a CCD. This can, in principle, reduce the noise below the 1/f barrier and deep into the sub-electron regime. As a proof of concept with our prototype SiSeRO device (Fig. \ref{fig:sisero}b), we employed clocking to the output gate and summing well to pull the charge back from the buried gate under the SiSeRO, carrying out nine repetitive transfers of the charge signal. The results are highly encouraging and suggest the potential for significant improvements in performance with future RNDR optimized SiSeRO devices.  

X-ray Charge Coupled Devices (CCDs)  \cite{Lesser15_ccd,gruner02_ccd,ccd_janesick01} have been the primary detector technology for soft X-ray instrumentation for more than three decades of X-ray astronomy. 
The latest generation of X-rays CCDs with advanced readout electronics exhibit impressive readout speeds ($\sim$2 MHz) and noise performance \cite{bautz18,bautz19,bautz20,chattopadhyay22_ccd,herrmann20_mcrc,Bautzetal2022,Oreletal2022}. 
However, the CCDs still lack in very high readout speed, region-of-interest (ROI) readout, and radiation hardness.
Other competing technologies, for example, Hybrid CMOS detectors (HCDs \cite{hull17,hull18_small_pixel,chattopadhyay18_HCDoverview}) and Monolithic CMOS (MCMOS \cite{kenter19_mcmos}) detectors, can provide very high readout speeds in tens of MHz range. 
These different detector technologies, however, either lack in low noise performance, in the case of HCDs, or sufficient detection efficiency beyond 5 keV in the case of MCMOS. SiSeRO with RNDR can provide very low noise readout (in sub-electron levels). In the future, RNDR optimized SiSeRO devices with a SiSeRO for every pixel, might remedy all of the aforementioned weaknesses of existing X-ray detection technologies, providing the capability to combine full frame, very low-noise readout with high speed, ROI readout in the same observation, while minimizing the sensitivity to radiation induced displacement damage. The RNDR technique also opens opportunities for unprecedentedly precise gain calibration \cite{rodrigues21}, where the instrument gain can be corrected in-situ by identifying individual electron bumps in the energy spectrum. This will be especially useful in characterizations in the low energy response of instruments ($<1$ keV), where much astrophysical discovery space resides \cite{reynolds23,gaskin19}.

In the next sections, we give a brief description of our experimental setup, the technical details of the  RNDR technique, and our experimental results. A summary and discussion of future plans can be found in Sec. \ref{sec:summary}.


\section{Repetitive non-destructive readout (RNDR) in SiSeROs}
\label{sec:rndr}
\subsection{Experimental setup} \label{subsec:setup}
\begin{figure}
    \centering
    \includegraphics[width=0.55\linewidth]{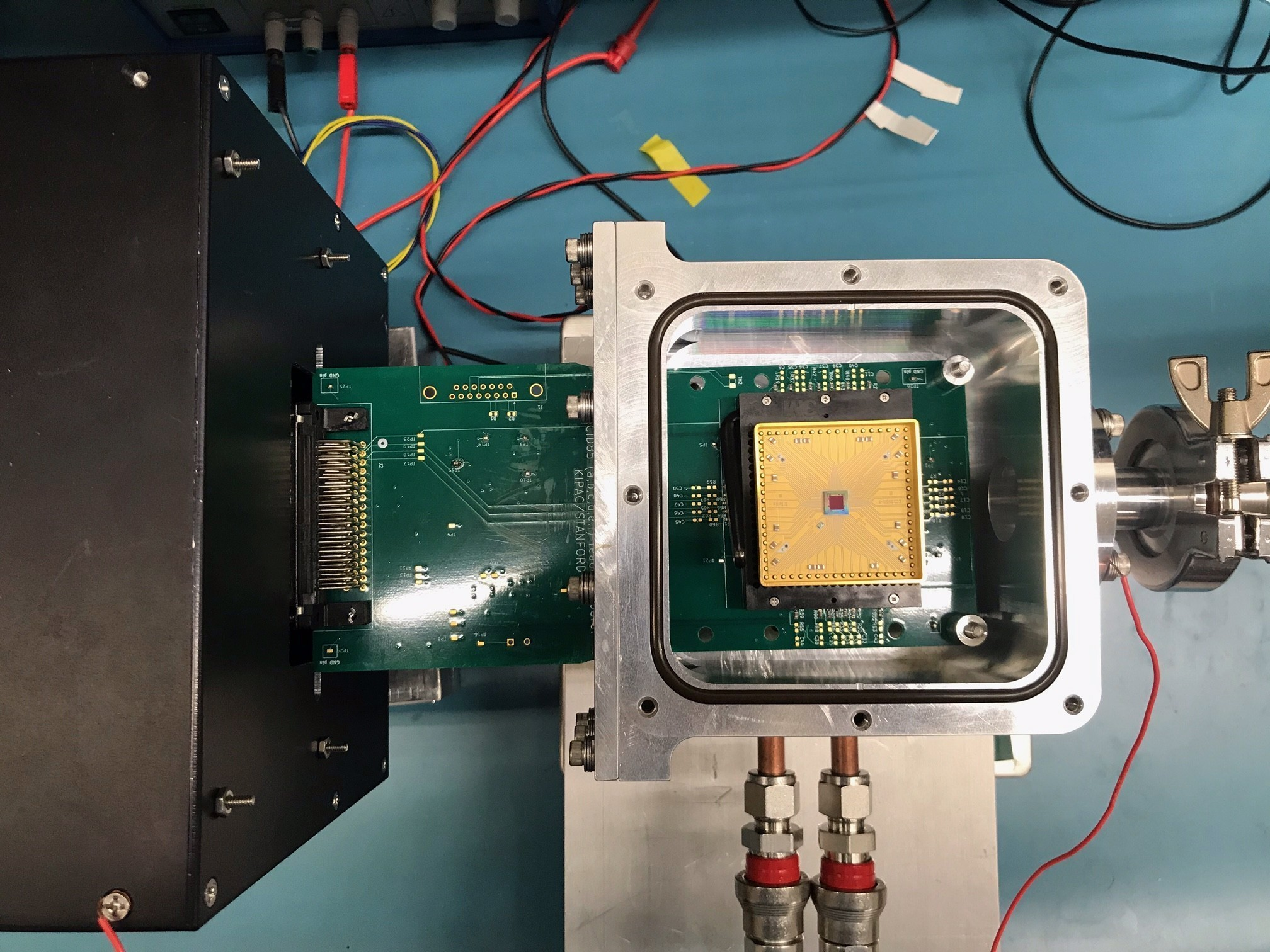}
    \caption{The experimental setup used for RNDR experiment. A CCID-93 buried channel SiSeRO device is mounted inside the chamber. A beryllium window mounted on the top flange (not shown here) is used for X-ray entrance. }
    \label{fig:setup}
\end{figure}

The experimental setup (also known as the `Tiny Box', see Chattopadhyay et al. 2020 \cite{chattopadhyay20_spie} for details) is shown in Fig. \ref{fig:setup}. 
A compact (13 cm $\times$ 15 cm $\times$ 6.5 cm) aluminum vacuum chamber houses the X-ray detector. The detector is mounted on an aluminum block, which is epoxied on a thermo-electric cooler (TEC) that is used to cool down the detectors. A liquid plate on the back of the bottom flange removes the heat deposited by the TEC. 
A Proportional-Integral-Derivative (PID) algorithm controls the detector temperature with better than 0.1$^\circ$C accuracy. A beryllium (Be) X-ray entrance window on the top flange allows X-ray photons to illuminate the detectors. A $^{55}$Fe radioactive source is placed on top of the Be window (outside the vacuum chamber) to characterize the spectral resolution of the detector.

As mentioned earlier, for this experiment, we used an MIT-LL CCID-93 prototype X-ray CCD with a buried-channel SiSeRO at the output stage. 
The readout electronics \cite{chattopadhyay22_sisero} consist of a custom preamplifier board (the circuit board in Fig. \ref{fig:setup}) and a commercial Archon controller \cite{archon14} (the black box in the figure). The preamplifier uses a drain current readout, where an I2V amplifier first converts and amplifies the SiSeRO output and a differential driver (at the second stage) converts the output to a fully differential signal.
The Archon, procured from Semiconductor Technology Associates, Inc (STA\footnote{\url{http://www.sta-inc.net/archon/}}), provides the required bias and clock signals to run the detector, and digitize the output signal.

\subsection{Repetitive non-destructive readout (RNDR) technique}
\label{subsec:rndr_technique}
Since in SiSeRO, the charge packet remains unaffected by the readout process (not transferred to a doped node as in conventional CCDs), we explored the possibility to utilize RNDR by moving the charge packet multiple times between the internal channel and the adjacent output gate. This technique has been demonstrated for DEPFET devices with sub-electron read noise yield \cite{wolfel06,treberspurg22_rndrdepfet}, and related efforts have been demonstrated on Skipper CCDs \cite{tiffenberg17}. An advantage of repetitive readout of the same charge signal is that the read noise can be reduced significantly, resulting in extremely low noise. It should also be noted that this technique keeps the full signal range intact, an important aspect for X-ray detectors where these ranges can span two orders of magnitude (0.1--10 keV). 

\begin{figure}
    \centering
   \hspace{-0.3cm}
    \includegraphics[width=\linewidth]{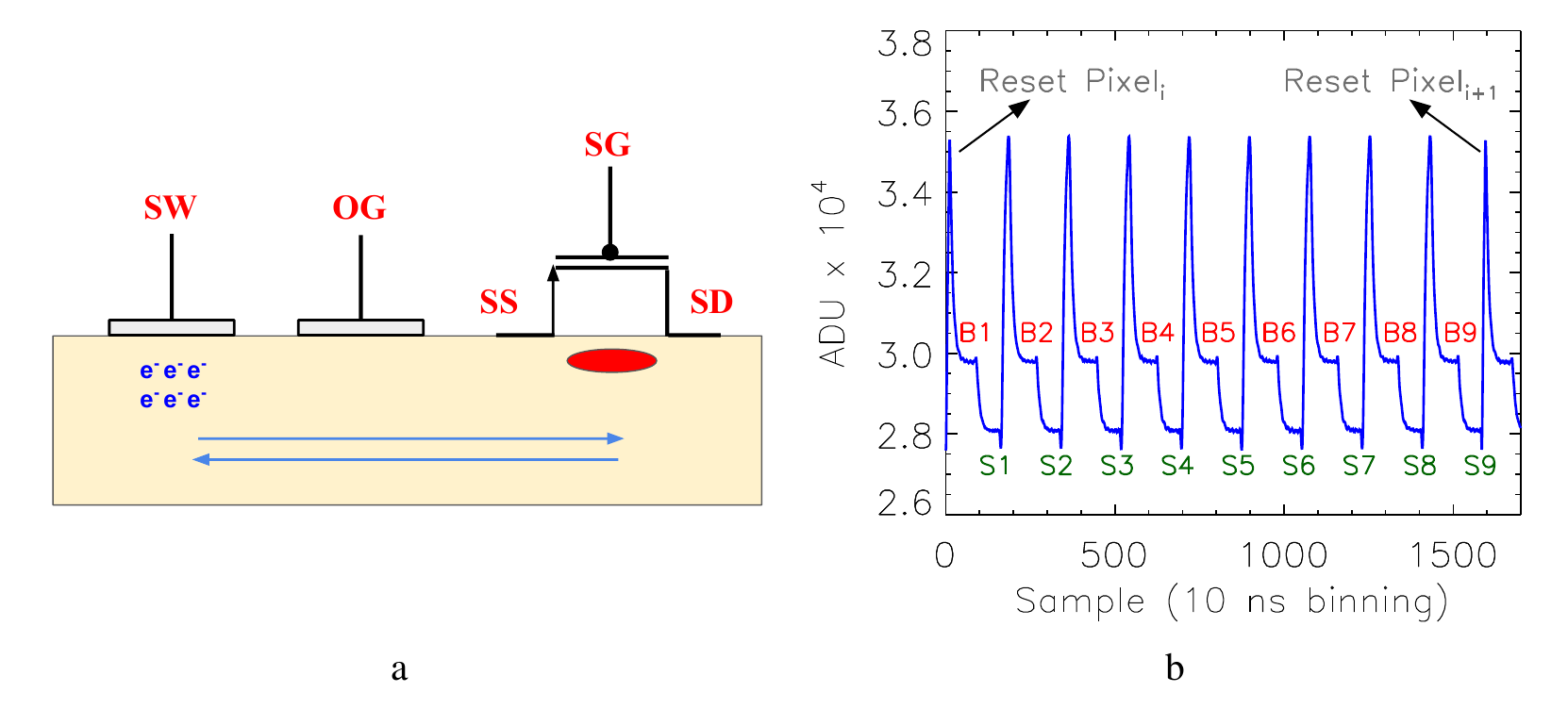}
    \caption{(a) Schematic of RNDR technique. The p-MOSFET SiSeRO transistor (SS: SiSeRO source, SD: SiSeRO drain, SG: SiSeRO gate) sit next to output gate (OG) and the last serial register gate (SW or summing well), allowing the signal charge to be moved between the buried back gate and SW for repeated measurement. (b) The waveform for one pixel over nine repetitive readout cycles showing baseline (B) and signal (S) measurements for each of the nine cycles.}
    \label{fig:rndr_method}
\end{figure}
In Fig. \ref{fig:rndr_method}, we show a simplified schematic of how RNDR is implemented in our setup, and the waveform for one pixel in RNDR readout. Here we implemented nine repetitive readout cycles, which increases the readout time by nine times overall (readout rate $\sim$63 kpixel/s). In the current setup, the detector temperature is limited to -23$^\circ$C. Therefore, the increase in readout time results in a larger thermal dark current contribution (which limits the repetitive readout to nine cycles in this experiment). For each repetitive cycle, we first move the charge from the internal channel back to the output gate (OG, see Fig. \ref{fig:rndr_method}a) by applying a positive clock potential ($>$ channel potential, which is around 2 V) to the OG. Note that in normal applications (non-RNDR), OG is connected to a positive DC bias (0.5 V). However, in RNDR, to move the charge back and forth, we clock OG between 0.5 V (OGLow) and 4 V (OGHigh). The last serial clock (Summing Well or SW in CCID-93) is then made high and OG is simultaneously made low, to move the charge to SW (Fig. \ref{fig:rndr_method}a). As OG settles down to OGLow, this introduces the individual baselines (from B$_2$ to B$_9$ in Fig. \ref{fig:rndr_method}b) that we see in the waveform. The charge packet is then moved back to the internal channel by changing the SW potential to SWLow ($<$OGLow). This introduces the individual signal regions (from S$_2$ to S$_9$ in Fig. \ref{fig:rndr_method}b) in the waveform. After the final readout, the reset transistor (see Fig. \ref{fig:sisero}a) is opened and a positive potential to the reset drain removes the charge from the internal channel, making the detector ready for charge transfer from the next pixel. 

\subsection{Repetitive non-destructive readout (RNDR) results} \label{subsec:rndr_results}
We first discuss the read noise measurements obtained using the RNDR technique. The read noise is calculated by measuring the width of distribution of signal in the overclocked pixels, which are an array of over-scanned pixels at the end of each pixel row. The amount of signal charge is expected to be negligibly small and the noise measured should be entirely due to the readout circuitry. 
\begin{figure}
    \centering
   \includegraphics[width=\linewidth]{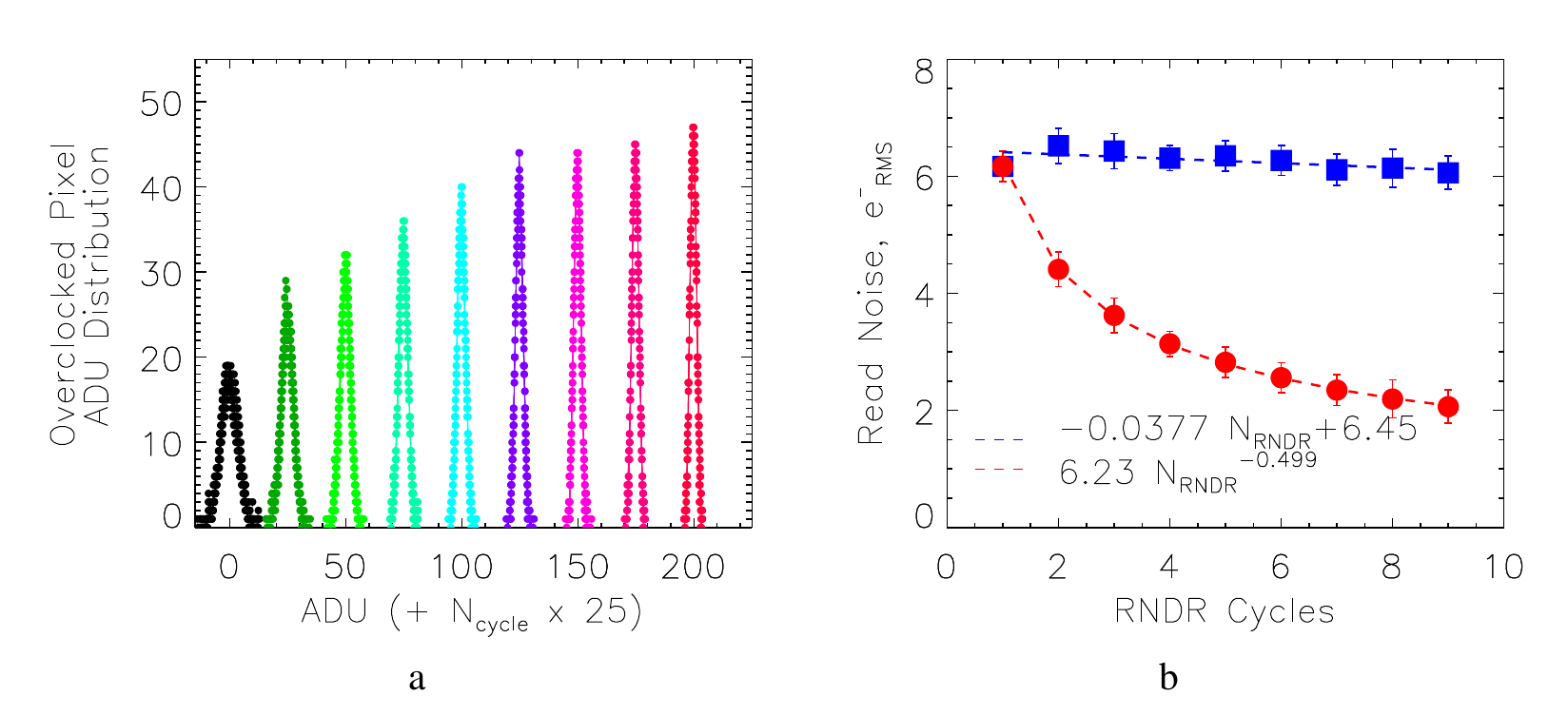}
    \caption{Read noise from the RNDR analysis. (a) Distribution of charge (in digital unit / ADU) obtained from the overclocked pixels. The distributions are plotted in an increasing order of the RNDR cycles from left to right. A constant offset of 25 ADU is added to the distributions for display purpose. With each RNDR cycle, the distribution achieves a narrower width. (b) The blue points show the individual read noise measurements from the 9 individual measurements while the red points are obtained when we average over consecutive measurements. The red data points follow the expected 1/$\sqrt{N}$ trend. The read noise for the first cycle is measured to be 6.17 $\mathrm{e}^{-}_{\mathrm{RMS}}$, while at the end of ninth cycle the read noise has improved to 2.06 $\mathrm{e}^{-}_{\mathrm{RMS}}$. The error bars are the 1$\sigma$ uncertainties on the noise measurements.}
    \label{fig:rndr_noise}
\end{figure}
In Fig. \ref{fig:rndr_noise}a, we show the charge distributions obtained after averaging over consecutive repetitive cycles. For example, the distribution shown in black (first from the left) is obtained at the end of the first cycle, while the distribution in red (rightmost) is obtained at the end of ninth cycle, after averaging over all nine measurements. The distributions all peak at zero ADU; however, for display purposes we add a constant offset of 25 ADU to the consecutive distributions. The distributions are then fitted with a Gaussian model to quantify the width of the distributions. 
In Fig. \ref{fig:rndr_noise}b, we show the consecutively averaged read noise measurements (red circles) along with the nine individual results (blue squares). The individual cycles yield similar noise measurements (shown by the blue dashed line fitted to the data), suggesting that the transfer of the signal charge into and out of the back gate of the transistor is sufficiently efficient. 
Note that there is a slight difference between the first and the second measurement. This stems from unequal baseline sampling for the first cycle. The first baseline follows the reset of the internal channel whereas for the remaining cycles, the baselines follow the OG clock pulse. The widths of the reset and OG pulses differ by some amount resulting in an unequal widths of the baseline regions.   
The red dashed line shows a fit with a power law model to the averaged measurements, ${N^\alpha_{\mathrm{RNDR}}}$, which yields a best fit value of -0.499 for $\alpha$. We achieved a read noise of 2.06 $\mathrm{e}^{-}_{\mathrm{RMS}}$ at the end of the ninth RNDR cycle. As mentioned earlier, we measure the responsivity of the device around 800 pA / electron which is close to the reported values for current DEPFET devices. Since this is a first generation SiSeRO device, we plan to improve the amplifier design in the next versions and thereby improve the responsivity of the amplifiers and consequently their noise performance.

\begin{figure}
    \centering
 \includegraphics[width=1\linewidth]{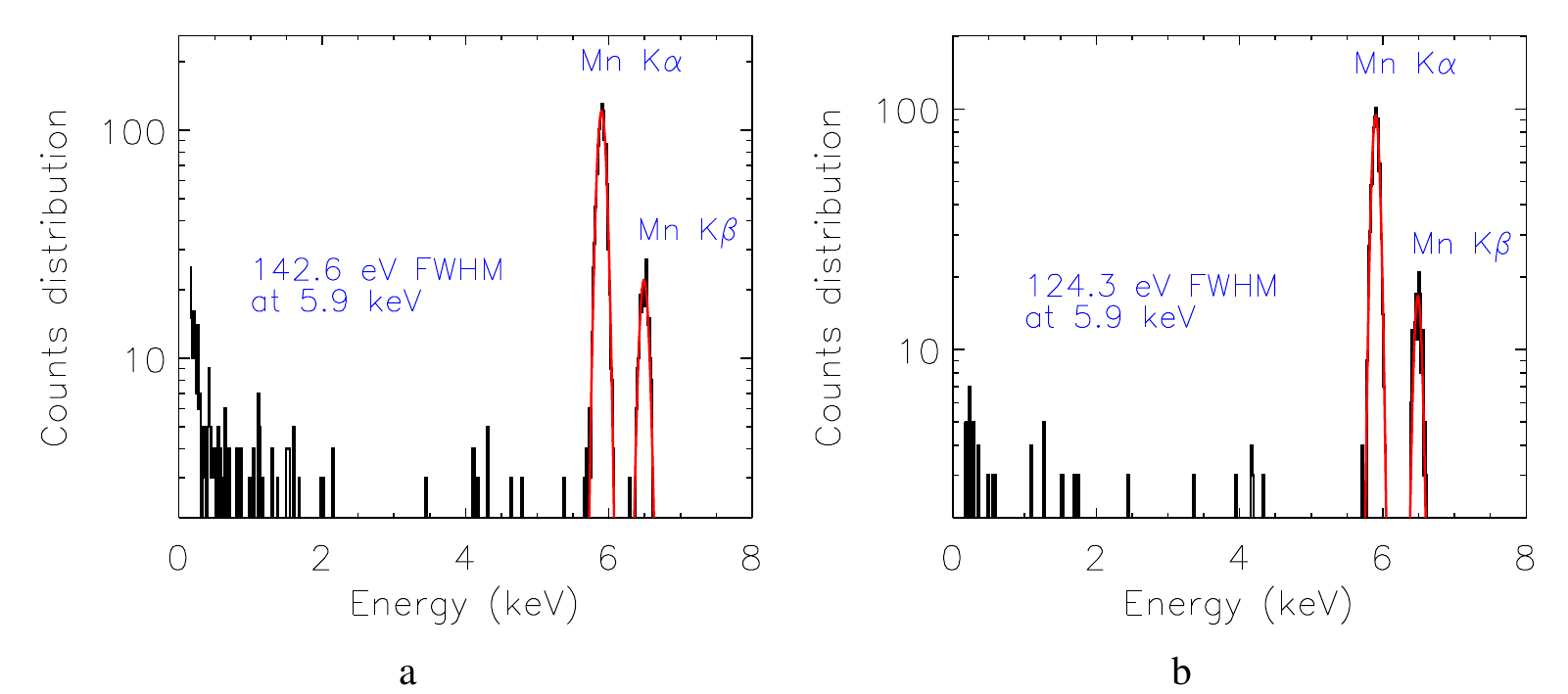}
    \caption{Spectrum obtained at the beginning (1$\mathrm{^{st}}$ RNDR cycle) (a) and at the end of the 9$\mathrm{^{th}}$ RNDR cycle (b) showing the Mn K$_\alpha$ (5.9 keV) and K$_\beta$ (6.4 keV) lines from a $^{55}$Fe radioactive source for single-pixel (grade 0) events. The measured FWHM at 5.9 keV improves from 143 eV to 124 eV. The Fano limit for silicon at 5.9 keV is 119 eV. The corresponding read noise for these two measurements are 6.17 and 2.06 $\mathrm{e}^{-}_{\mathrm{RMS}}$ respectively.}
    \label{fig:spectra}
\end{figure}

We used a $^{55}$Fe radioactive source to measure the spectral resolution of the device. The resulting spectra of Mn K$_\alpha$ (5.9 keV) and K$_\beta$ (6.4 keV) lines generated from the single-pixel events are shown in Fig. \ref{fig:spectra} for the 1$\mathrm{^{st}}$ and 9$\mathrm{^{th}}$ RNDR cycles. 
To generate this spectrum, the X-ray images were corrected for bias, by subtracting the overclocked region for each frame, and then for dark current, by subtracting dark current obtained after averaging over dark frames. An event list was generated from the corrected image frames, with each event consisting of signal amplitudes of 9-pixels around every local maximum in spatial distribution (presumably due to X-ray interaction with silicon). The selection of such events is based on a primary threshold (7 times the read noise). We apply a secondary threshold (2.6 times the read noise) to determine whether an event contains signal charge in the pixels surrounding the center and grade the events depending on number of pixels containing such extra charge. Spectra for each of the event grades are generated by adding charge in the adjacent pixels exceeding the secondary threshold.
The FWHM at 5.9 keV at the end of 9$\mathrm{^{th}}$ RNDR cycle is measured to be around 124 eV, whereas the single read measurement is around 143 eV. Note that the Fano limit for silicon at 5.9 keV is $\sim$119 eV.  

\begin{figure}
    \centering
    \includegraphics[width=\linewidth]{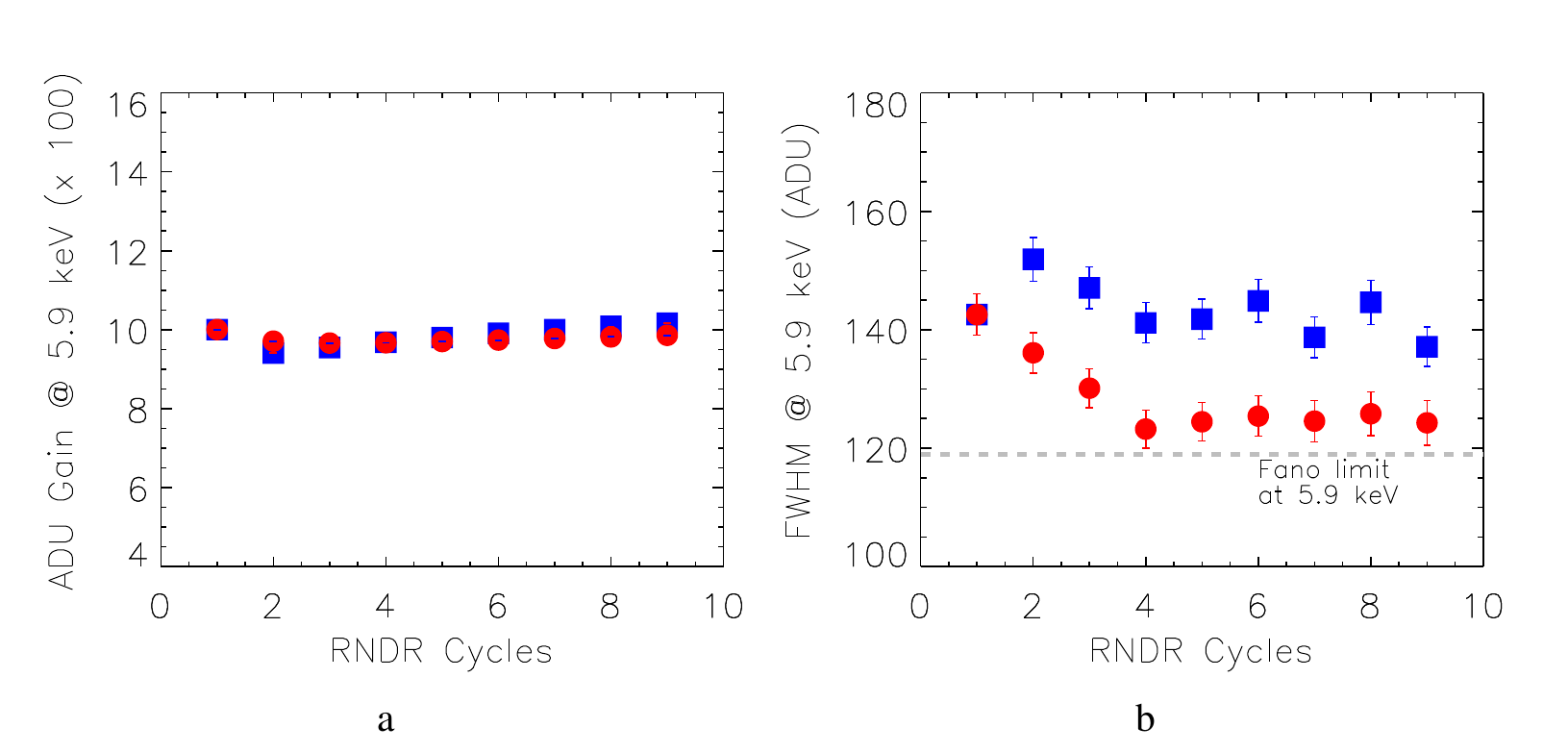}
    \caption{(a) Measured gain in digital unit / ADU at 5.9 keV as a function of RNDR cycles. The blue points show the individual gains for the 9 individual measurements, while the red points are obtained after averaging over consecutive measurements. The gains remain constant, implying no charge loss during the repetitive transfer of charge. (b) Measured FWHM at 5.9 keV as a function of RNDR cycles. The individual measurements (blue data points) lie within the $\sim$1$\sigma$ error bar of each other. The measurements obtained from the RNDR cycles (red) show a clear improvement in spectral resolution with the results approaching fano limit (119 eV) for these energies in silicon. With thermmal dark current dominating the total noise (ddark noise noise $\gg$ read noise), the FWHM measurements do not improve after the fourth cycle.}
    \label{fig:gain_fwhm}
\end{figure}

In order to examine the charge transfer efficiency with each transfer cycle, we compare the measured gain in digital unit / ADU at 5.9 keV as a function of RNDR cycles (see Fig. \ref{fig:gain_fwhm}a). 
The blue points show the individual gain measurements from the 9 individual measurements while the red points are obtained when we average over consecutive measurements. The gains remain fairly constant implying no charge loss during the nine repetitive transfer of charge. However, in order to quantify the charge transfer efficiency (CTE) or charge loss during the repetitive transfers, we require a larger number of repetitive cycles which we plan to do in future.
As mentioned earlier, there is a slight difference in the baseline region in the first transfer cycle from the rest, giving rise to slight change in the gain for the first measurement (and therefore the read noise). The measured FWHMs at 5.9 keV as a function of RNDR cycles are shown in Fig. \ref{fig:gain_fwhm}b. The individual measurements (blue data points) all lie within the $\sim$1$\sigma$ error bar of each other. The measurements obtained from the RNDR cycles (red) show a clear improvement in the spectral resolution with this method, approaching the Fano limit (119 eV) for these energies in silicon. With thermal current dominating the total noise after the fourth RNDR cycle, we do not see any further improvement in FWHM. We note that the read noise is estimated from the overclocked pixel array, which has a negligible thermal dark current contribution and therefore does not affect the read noise measurements. In future work, with an advanced setup providing better control of the detector temperature, the thermal dark current should be negligibly low, while keeping the number of RNDR cycles high.   


\section{Summary and future plans}\label{sec:summary}

The SiSeRO amplifier, developed by MIT Lincoln Laboratory, is a novel technology for the output stage of X-ray CCDs. 
The charge packet remains unaffected in the readout process, which offers the possibility to transfer the charge non-destructively between the output stage amplifier and the transfer gates, known as RNDR. At the end of each cycle, the same charge packet can be measured and, over many such measurements, the final average noise can be reduced by a factor of $\mathrm{1/\sqrt{N_{cycle}}}$.   

In this work, we implemented the RNDR technique with nine repetitive cycles on a prototype SiSeRO device, obtaining 2.06 $\mathrm{e^{-}_{RMS}}$ read noise at the end of the ninth cycle. This represents a factor of 3 (square root of 9) improvement for the individual measurement of around 6 $\mathrm{e^{-}_{RMS}}$ noise. The individual cycles are readout at 625 KHz speed over the nine cycles, for an effective readout speed of $\sim$63 KHz. Due to the limitations of our current setup, we limited the repetitive readout to nine cycles in order to keep the accumulated thermmal dark current low.  

Significant improvement in noise performance, potentially to sub-electron levels, can in principle be achieved by increasing the number of repetitive cycles. In future work with RNDR-optimized SiSeRO devices, we plan to achieve these goals with the following upgrades in our setup, readout electronics and detector output stage: 
\begin{itemize}
    \item A new X-ray test setup, currently under development, will support operation at temperatures down to -100$^\circ$C. This should reduce the thermal dark current noise and allow for a larger number of RNDR cycles.
    \item We have developed an ASIC-based system at Stanford to readout these devices (Herrmann et al. 2020 \cite{herrmann20_mcrc} and Orel et al. 2022 \cite{Oreletal2022}). The ASIC is expected to provide higher readout speed and lower noise compared to the existing discrete electronics. With these changes, it should be possible to achieve noise as low as 1 $\mathrm{e^{-}_{RMS}}$ or better, even with first generation SiSeRO devices.  
    \item For our proof of principle measurements, we employed a simple clocking of the output gate and summing well to pull the charge back from the buried gate under the SiSeRO. A potential improvement is to utilize two SiSeRO p-MOSFETs next to each other, allowing the charge packet to be transferred between the two. Such a design would allow for the fast transfer of the signal charge between the transistors and higher operational RNDR switch frequencies. This concept is comparable to that of RNDR optimized DEPFET detectors \cite{wolfel06}. We will conduct device simulations with the MIT Lincoln Laboratory process development kit to investigate various arrangements and layouts. Our SiSeRO building block can be the basis for other RNDR devices: for example, a cluster of multiple SiSeRO stages at the output should allow us to perform additional measurements of the same signal charge, resulting in a reduction of noise by an additional 1/$\mathrm{\sqrt{N}}$ while maintaining a high frame rate. 
    One of our long-term goals is to build SiSeRO active pixel sensor (APS) arrays in which each pixel will include the basic SiSeRO building block. This will allow extremely low noise performance at high readout speeds. In our preliminary layout design, we could accommodate two-SiSeRO structures in 24 $\mu$m pixel size. In the first devices, we plan to include 16 $\times$ 16 pixels, where the readout can be done using our existing 8 channel ASIC wire-bonded to one side of the matrix, while matrix control can be performed with a separate ASIC on the other side.       
\end{itemize}

SiSeROs with RNDR offer an exciting solution for future low-noise spectroscopic imagers for next generation astronomical X-ray telescopes. In addition, RNDR optimized SiSeROs provide opportunities for the precise gain calibration of these detectors  \cite{rodrigues21}, allowing to conduct in-situ absolute calibration at low X-ray energies ($\mathrm{<}$1 keV). In combination, these characteristics should help open up a new discovery space for X-ray astronomy.


\subsection*{Code, Data, and Materials Availability}
The data presented in this article are publicly available in ``jatis-2023-sisero-rndr" at \url{https://github.com/kipac-xoc/jatis-2023-sisero-rndr}

\subsection* {Acknowledgments}
This work has been supported by NASA grants APRA 80NSSC19K0499 ``Development
of Integrated Readout Electronics for Next Generation X-ray CCDs” and SAT
80NSSC20K0401 ``Toward Fast, Low-Noise, Radiation-Tolerant X-ray Imaging Arrays for
Lynx: Raising Technology Readiness Further.”








\end{spacing}
\end{document}